\documentclass[aps,prb,twocolumn,showpacs,superscriptaddress,amsmath,amssymb]{revtex4-1}

\usepackage{xcolor}
\usepackage{graphicx}

\title{\textbf{Dimensionality effects on the luminescence properties of hBN}}

\begin{document}

\title{Dimensionality effects on the luminescence properties of hBN}
\author{L\'eonard Schu\'e}
\affiliation{Laboratoire d'Etude des Microstructures, ONERA-CNRS, Universit\'e Paris-Saclay, BP 72, 92322 Ch\^atillon Cedex, France}
\affiliation{Groupe d'Etude de la Mati\`ere Condens\'ee, UVSQ-CNRS, Universit\'e Paris-Saclay, 45 avenue des Etats-Unis, 78035 Versailles Cedex, France}
\author{Bruno Berini}
\affiliation{Groupe d'Etude de la Mati\`ere Condens\'ee, UVSQ-CNRS, Universit\'e Paris-Saclay, 45 avenue des Etats-Unis, 78035 Versailles Cedex, France}
\author{Andreas C. Betz}
\affiliation{Laboratoire Pierre Aigrain, Ecole Normale Sup\'erieure-PSL Research University, CNRS, Universit\'e Pierre et Marie Curie-Sorbonne Universit\'es, Universit\'e Paris Diderot-Sorbonne Paris Cit\'e, 24 rue Lhomond, 75231 Paris Cedex 05, France}
\affiliation{Hitachi Cambridge Laboratory, JJ Thomson Avenue, CB3 0HE Cambridge, United Kingdom}
\author{Bernard Pla\c{c}ais}
\affiliation{Laboratoire Pierre Aigrain, Ecole Normale Sup\'erieure-PSL Research University, CNRS, Universit\'e Pierre et Marie Curie-Sorbonne Universit\'es, Universit\'e Paris Diderot-Sorbonne Paris Cit\'e, 24 rue Lhomond, 75231 Paris Cedex 05, France}
\author{Fran\c{c}ois Ducastelle}
\affiliation{Laboratoire d'Etude des Microstructures, ONERA-CNRS, Universit\'e Paris-Saclay, BP 72, 92322 Ch\^atillon Cedex, France}
\author{Julien Barjon}
\affiliation{Groupe d'Etude de la Mati\`ere Condens\'ee, UVSQ-CNRS,  Universit\'e Paris-Saclay, 45 avenue des Etats-Unis, 78035 Versailles Cedex, France}
\author{Annick Loiseau}
\affiliation{Laboratoire d'Etude des Microstructures, ONERA-CNRS, Universit\'e Paris-Saclay, BP 72, 92322 Ch\^atillon Cedex, France}
\email{annick.loiseau@onera.fr}

\begin{abstract}
Cathodoluminescence (CL) experiments at low temperature have been undertaken on various bulk and exfoliated hexagonal boron nitride (hBN) samples. Different bulk crystals grown from different synthesis methods have been studied. All of them present the same so-called S series in the 5.6--6~eV range, proving its intrinsic character. Luminescence spectra of flakes containing 100 down to 6 layers have been recorded. Strong modifications in the same UV range are observed and discussed within the general framework of 2D exciton properties in lamellar crystals.
\end{abstract}


\maketitle

\section{Introduction}

Black and white graphenes are one-atom thick layers of graphite and hexagonal boron nitride (hBN), respectively. Their remarkable properties are inherited from their two-dimensional (2D) crystal structure and symmetry: Graphene is a semi-metal populated by massless chiral Dirac fermions while hBN is a large band gap semiconductor ($>$ 6 eV).  As an ultimate 2D crystal, single-layer graphene (SLG) is highly sensitive to its environment and its unique properties can easily be spoiled, but may also be enhanced by the close proximity of a substrate. Therefore, having mostly understood the potential of intrinsic graphene, the new challenge is to engineer its coupling to substrates \cite{Ju2014,Woods2014} and exploit it to realize innovative electronic and optoelectronic devices. It has become quite clear in recent years that the most compatible environment for graphene is hBN, due to both its insulating character and its similar honeycomb crystal structure, which matches almost perfectly that of graphene.\cite{Dean2010,Wang2013,Mishchenko2014} Further graphene and hBN have excellent material characteristics, in particular a strong $sp^{2}$ bonding that sets a robust energy scale for all excitations (electrons, phonons, excitons). 

The rise in widely using hBN layers as a substrate or encapsulating layers of graphene is mainly based on their ability to preserve at the best electronic properties of graphene such as the carrier mobility. \cite{Dean2010} However, it was recently shown that, when graphene is epitaxially stacked on hBN layers, a strong interlayer coupling can arise, resulting in marked modifications of its electronic properties. \cite{Yankowitz2012,Yang2013} An appropriate control of the quality and properties of hBN layers, which may be in turn in an interlayer coupling seems, therefore, an essential milestone towards an appropriate use of hBN layers in complex heterostructure architectures and devices.  

However, in contrast to graphene, electronic properties of BN materials remain basically to understand. Luminescence spectroscopies are here very useful tools. Dedicated cathodoluminescence (CL) and photoluminescence (PL) experiments at 10K have been recently developed and used to study BN powders, single crystals and nanotubes.\cite{Watanabe2004,Watanabe2006,Watanabe2006b,Silly2007,Jaffrennou2008,Pierret2014,Pierret2015}

Luminescence of hBN is found to be dominated by near band edge excitonic recombinations, described as S and D lines. The  higher-energy S lines, between 5.6 and 5.9 eV, are generally attributed to intrinsic free excitons, whereas the lower-energy D ones, between 5.4 and 5.65 eV, are assigned to excitons trapped in structural defects.\cite{Jaffrennou2007,Watanabe2009} Very little is known on hBN 2D layers. In 2D crystals, the excitonic effects are generally amplified by the spatial confinement of electrons and holes, and by the decrease of the electrostatic screening. Lamellar crystals have totally original features when compared to standard 3D-semiconductors, even in the confinement regime reached with epitaxial quantum wells. For instance in MoS$_{2}$, the band structure evolves from an indirect bandgap in bulk crystals to a direct band gap for the single atomic layer \cite{Mak2010} having a 1 eV exciton binding energy.\cite{Qiu2013} More recently, a similar indirect-to-direct gap crossover has been reported in WS$_{2}$ and WSe$_{2}$ from few-layer flakes to the single-layer.\cite{Zhao2012} Analogous effects are then expected for hBN for which exciton binding energies up to 2.1 eV have been theoretically predicted for the single layer.\cite{Arnaud2006,Arnaud2008,Wirtz2005,Wirtz2006,Wirtz2008}

Here, we present a study of the luminescence properties of hBN from its bulk form to atomic exfoliated thin flakes. A thorough investigation  by cathodoluminescence (CL) at 10 K of different hBN bulk sources is used to highlight the intrinsic origin of the S series. On the basis of our previous experiments,\cite{Pierret2014} we manage to isolate almost defect-free exfoliated flakes of different thicknesses from 100 ML down to a 6 ML and to explore their intrinsic luminescence by cathodoluminescence. Measurements reveal a strong thickness-dependent evolution of the near band edge emission that provides insight into the nature of the excitonic recombinations in both the bulk and the 2D layers.

\section{Experiments}

The bulk hBN materials investigated in this work are from different origins and were synthesized by various growth processes. Commercially available, the Saint-Gobain powder, dedicated to cosmetic applications (Tr\`esBN\textsuperscript{\textregistered}PUHP1108), was synthesized at high temperature from boric acid and nitrogen source. BN crystallites from HQ-graphene company were also examined (HQ sample). A sample provided by the Laboratoire Multimat\'eriaux et Interfaces (LMI sample) was synthesized following the Polymer Derived Ceramics (PDCs) route utilizing a polymeric precursor converted at high temperature to a ceramic.\cite{Yuan2014,Yuan2014b} Finally a single crystal grown at high-pressure high-temperature (HPHT) provided by the NIMS \cite{Taniguchi2007} was taken as a reference.

The BN sheets studied in this paper were all exfoliated from Saint-Gobain crystallites. Mechanical peeling was used following the method initially developed for graphene.\cite{Novoselov2004} The powder is applied to an adhesive tape, whose repeated folding and peeling apart separates the layers. They are further transferred onto a Si wafer covered with 90 nm of SiO$_{2}$, the optimal thickness for imaging BN flakes with a maximal optical contrast. \cite{Gorbachev2011} The SiO$_{2}$/Si substrates are also covered with a network of Cr/Au finder marks deposited by UV lithography with the AZ5214E photoresist and Joule evaporation in order to facilitate the localization of the flakes.  Prior to  the layer transfer, the wafer was chemically cleaned with acetone and isopropanol, followed by several minutes of exposure to a O$_{2}$ plasma (60W, P$\leqslant$12 nbar). The last step of this preparation renders the SiO$_{2}$ surface hydrophilic and more sensitive to water contamination. That is the reason why we prefer to use the optical contrast (OC) for thickness determination after a Atomic Force Microscopy (AFM) calibration procedure on folded parts (see description in Supporting Information).

The optical properties of BN samples were analyzed by cathodoluminescence using an optical system (Horiba Jobin Yvon SA) installed on a JEOL7001F field-emission gun scanning electron microscope (SEM). The samples are mounted on a GATAN cryostat SEM-stage cooled down to 5 K with a continuous flow of liquid helium. The samples are excited by electrons accelerated at 2 kV with a beam current of 1 nA. Exfoliated layers were measured using either a fixed electron beam excitation, or for the thinnest flakes, a fast e-beam scanning on the sample in order to limit e-beam-induced modifications. The CL emission is collected by a parabolic mirror and focused with mirror optics on the entrance slit of a 55 cm-focal length monochromator. The all-mirror optics combined with a suited choice of UV detectors and gratings ensures a high spectral sensitivity down to 190 nm. A silicon charge-coupled-display (CCD) camera is used to record the spectra. The spectral response of the optical detection setup was measured from 200 to 400 nm using a deuterium lamp (LOT Oriel - Deuterium lamp DO544J - 30W) of  calibrated spectral irradiance. The system response for each detector/grating combination was then obtained following the procedure described in Ref.\citenum{Pagel2000}. All spectra reported in this paper are corrected from the system response. It was also checked that the CL linewidths are not limited by the spectral resolution of the apparatus (0.02 nm in the best conditions) excepted for the 6L flake acquisitions.

\section{Results}

\subsection{Intrinsic luminescence of bulk hBN}

\begin{figure}[ht]
 \begin{center}
\includegraphics[scale=0.8]{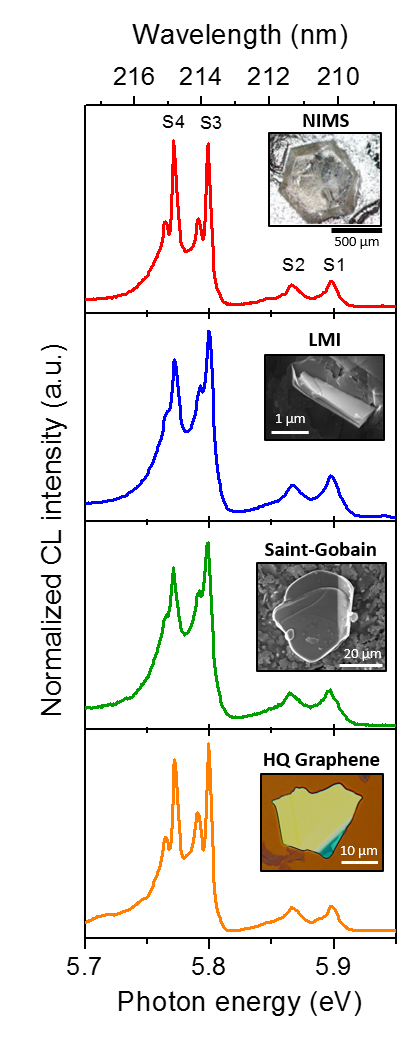}
\caption{CL corrected spectra of hBN bulk hBN materials in the near-band-edge region. The reference spectrum of a HPHT high quality single crystal from NIMS is compared to the chemically-grown LMI sample PDCs and the commercial hBN samples. The sample holder temperature is was 5K. All spectra are corrected from the spectral response of the detection system. In insert, typical images of the hBN materials are shown.}
\label{CLhbn}
 \end{center}
 \end{figure}

The CL spectra taken from the commercial samples (Saint-Gobain and HQ Graphene), the LMI sample and the reference crystal from NIMS are reported together in Fig.\ \ref{CLhbn}. Actually as these samples were synthesized using very different growth techniques whether it is concerning  boron and nitrogen precursors, pressure  or process temperature, they obviously cannot afford identical defects or impurities, which could act as trapping sites of the excitons. \cite{Pierret2014,Jaffrennou2007} Nevertheless, in spite of their differences, for all the samples, we observe a similar  S series emission with the S1, S2, S3 and S4 lines emerging with a maximum at 5.898 eV (210.2 nm), 5.866 eV (211.4 nm), 5.797 eV (213.9 nm) and 5.771 eV (214.8 nm) respectively, as already reported in previous works. \cite{Watanabe2009,Cao2013,Pierret2014}  This comparative study therefore strengthens the current interpretation attributing an intrinsic origin to the S lines. \cite{Jaffrennou2007,Watanabe2009} \\
According to this interpretation, the S emission lines observed in CL and PL have been assigned \cite{Jaffrennou2007,Watanabe2009} to the theoretical excitons calculated by Arnaud and Wirtz. \cite{Arnaud2006,Arnaud2008,Wirtz2006,Wirtz2008,Wirtz2009} Actually these calculations show that in the case of a perfect hexagonal symmetry, hBN present two doubly degenerate exciton levels, with a lower dark pair and an upper bright one. A symmetry lowering lifts the degeneracies so that four non degenerate levels are obtained.
\footnote{There are also dark triplet states which are not discussed here.}
It is not known precisely what is the reason for this symmetry lowering in our case: zero-point motion\cite{Arnaud2008} or Jahn-Teller like effect\cite{Watanabe2009} have been suggested. In any case phonon effects and exciton-phonon couplings should be involved as indicated by the Stokes shift determined from the comparison of absorption measurements with the PL-CL spectra.\cite{Watanabe2011} It should be noticed that these interpretations assume that the main relevant absorption and luminescence processes are related to direct transitions although {\it ab initio} calculations predict an indirect gap between points in the Brillouin zone close to the K (and H) point for the valence and and close to the M (and L) point for the conduction band. One- or multiphonon processes should perhaps be taken into account.\footnote{G. Cassabois, P. Valvin and B. Gil, arXiv:1512.02962 [cond-mat.mtrl-sci]}

Finally it is worth mentioning that the excitons described here present similarities with those observed in lamellar transition metal dichalcogenides (TMDs) in which case spin-orbit splitting is clearly observed. \cite{Qiu2013,Xu2013,Berghauser2014,Zhang2015} However the spin-orbit splitting should be very weak here for the light B and N elements. Looking at the spectra in more detail, the S3 and S4 emissions exhibit a fine structure splitting with a doublet separated by a few meV and assigned to the transverse and longitudinal components of exciton recombinations, as suggested by polarized-PL experiments.\cite{Watanabe2009} The S3 and S4 doublets are observed leading to the two S4 peaks at 5.762 eV and 5.768 eV and the two S3 peaks at 5.787 eV and 5.794 eV. They can be distinguished in the LMI and the Saint-Gobain powder samples but are better seen in the HQ and the NIMS crystals, the latter generally exhibiting narrower linewidths. We believe that this is due to the fact that the HQ and the NIMS samples were cooler, because mounted on a copper plate, than the two other ones, mounted on SiO$_{2}$/Si substrates of lower thermal conductivity. Indeed, the S peak linewidths have been shown to be strongly dependent on the crystal temperature.\cite{Watanabe2009}

\subsection{Thickness and luminescence of hBN flakes}

Several examples of exfoliated flakes are presented in Fig.\ \ref{afm} and in Supplementary Information. The uncommon geometry of the exfoliated flake shown in Fig.\ \ref{afm} offers a good configuration to trace the luminescence properties as a function of the hBN thickness in the nanometer range. It displays three atomically-flat regions of 9, 20 and 32 nm thickness as deduced from the AFM profile (Fig.\ \ref{afm}b). Assuming a hBN interplanar distance of 0.34 nm and a 1 nm water layer thickness, we estimate that the exfoliated sample is composed of 27, 60 and 100 layers respectively. It is important to notice that the contour outline of the flake is well-defined, suggesting that the hBN crystal is free of any glide defect, which is also confirmed by the absence of the D  emission series in the CL spectra.\cite{Pierret2014,Watanabe2009} This strongly suggests the pristine AA' stacking of hBN in these three thicknesses. The thickness steps are probably produced by a cleavage perpendicular to the basal plane during the tape removal. Planes with lowest indices (10.0) and (11.0) being the easier to split, the parallel step edges are thus probably oriented along the [11.0] and [10.0] directions respectively in the basal plane corresponding to armchair and zigzag edges respectively. Actually, a recent study reported a statistic of the edges structure in hBN exfoliated flakes and showed a slight preference for zigzag-type edges.\cite{Kim2011}

\begin{figure}[ht]
\begin{center}
\includegraphics[scale=0.6]{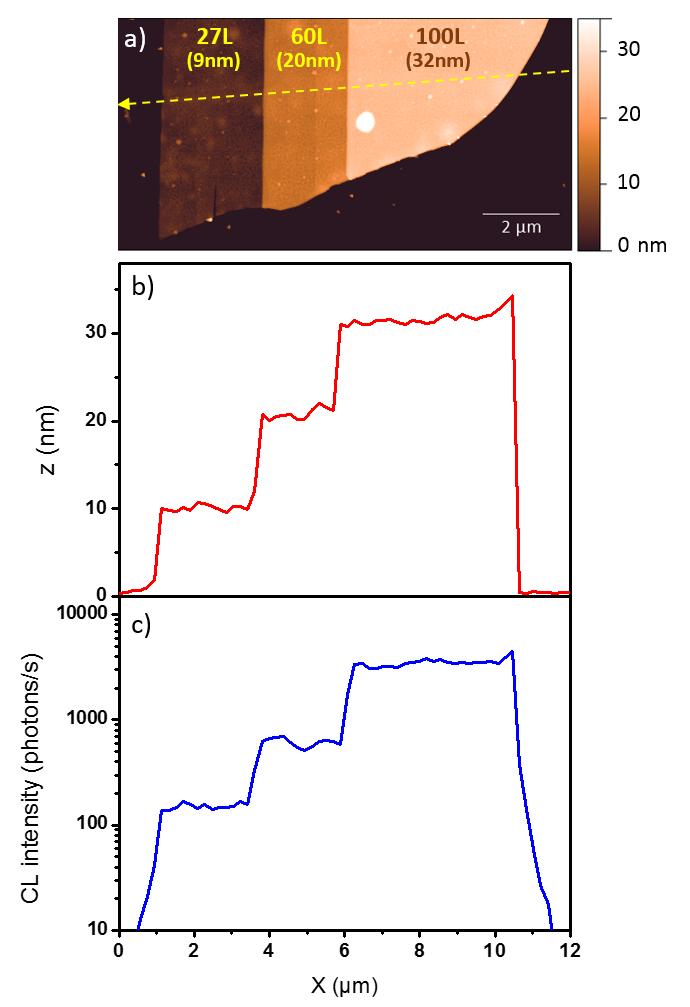}
\caption{a) AFM image of a mechanically exfoliated hBN flake with the dotted arrow along which was recorded the profile depicted in b) showing three distinct height steps of 9, 20 and 32 nm respectively. The profile of integrated CL signal of the S lines is plotted in c).}\label{afm}

\end{center}
\end{figure}

A CL linescan recorded perpendicularly to the step edges is reported in Fig.\ \ref{afm}c. The CL intensity clearly decreases when reducing the hBN thickness. Such a result is expected because the electron-hole generation rate decreases with the thickness, thin flakes being almost transparent to the electron beam. However, it is remarkable that we obtain an almost homogeneous intensity of S emissions for a given thickness, also consistent with the integrity of the hBN crystal after exfoliation.

\begin{figure}[ht]
\begin{center}
\includegraphics[scale=0.8]{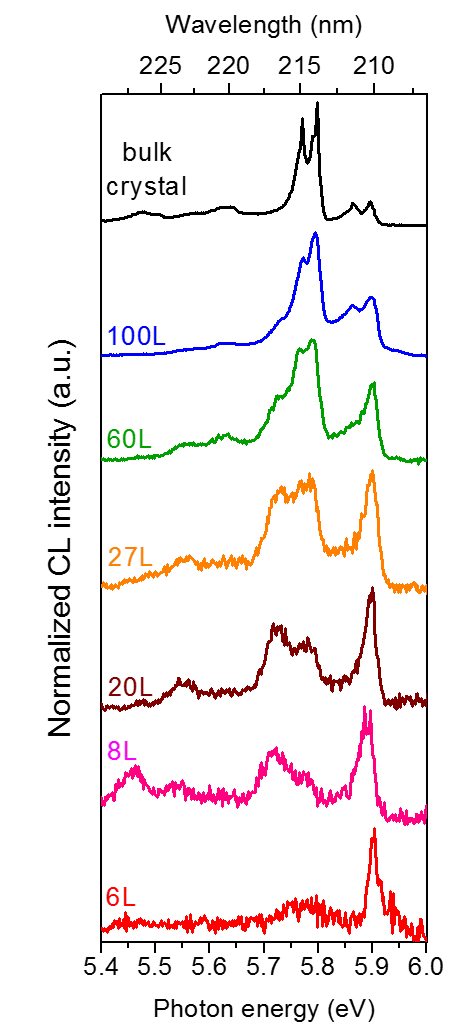}
\caption{Normalized CL spectra of bulk hBN and flakes of 100, 60, 27, 20, 8 and 6L obtained from the Saint-Gobain material. All spectra were corrected from the spectral response of the detection system (details in the text). The sample holder temperature was 5K.}
\label{super}
\end{center}
\end{figure}

We stress here that we managed to isolate hBN layers free of exfoliation defects, which was not the case in our previous work. \cite{Pierret2014} It is attested by the CL spectra being now all dominated by the S exciton recombinations. The trend extracted from the defect-free Saint-Gobain flakes is summarized in Fig.\ \ref{super} showing the CL spectra for thicknesses ranging from 100L to 6L. The spectrum of the bulk Saint-Gobain crystallite is also plotted for comparison. Strong modifications in the CL spectra appear within the 5.6-6 eV energy range for exfoliated layers. When reducing the thickness, one can see a significant decrease of the relative intensities of the S4, S3 and S2 lines compared to the S1 line. At the level of a few monolayers, it is remarkable that the characteristic S4, S3 and S2 lines of bulk hBN almost completely vanished.

This low dimensionality effect is not sample dependent and was in particular confirmed with hBN flakes exfoliated from the NIMS sample (see SFig.2 of the Supporting Information) leading to the conclusion that the observed thickness dependence is not governed by defect generation but has rather an intrinsic behavior. The S1 emission, which dominates the luminescence spectra of hBN flakes thinner than 20L, remains particularly sharp. The FWHM of this peak for the 6L sample is (17meV-0.6nm) which corresponds to the spectral resolution of the used experimental conditions. Besides, we observed a small energy shift of the S1 line from 5.871 eV for the bulk to 5.909 eV for the thinnest 6L flake. This shift could be a first indication of a modification in the intrinsic exciton emission energy such as a change in bandgap and/or in exciton binding energy. A similar interpretation has been proposed in a recent study wherein this singular emission at 5.9eV was observed in large-diameter BNNTs (60nm).\cite{Du2014}

\section{Discussion}

The only available {\it ab initio} study comparing bulk hBN and monolayer properties modeled the effect of varying the interplane distance.\cite{Wirtz2006} The calculated absorption spectra including excitonic effects shows that the bandgap increases with confinement, but the exciton binding energy also does in the same proportion, so that absorption peaks are expected to be in the same range of energy for the bulk and the monolayer ($+0.2$ eV). Given the uncertainty of such calculations, they appear consistent with our observations of a few tens of meV energy shifts of the S lines when reducing the number of layers of hBN.

Looking closer to the spectra for the 90L flake, a shoulder on the low energy side of the S4 peak can also be detected. For thinner samples, it emerges --- together with the S1 line --- as a broader peak of about 50 meV linewidth having a maximum at 5.726 eV (216.5 nm). This peak may be assigned to the LO phonon (169 meV) replica of the S1 line (5.898 eV). This assignment is strengthened by the observation of replica involving up to 3 phonons when using higher integration times (see Fig.S3 of the Supporting Information).

Flakes composed of less than 6L were also investigated in this work on SiO$_{2}$/Si substrates but we have not yet succeeded in getting a significant CL signal in the 1-5 layers thickness range. Regarding the luminescence efficiency, the situation of BN on SiO$_{2}$ is in contrast to that of suspended MoS$_{2}$ layers where the quantum yield increases by a factor of 10$^{4}$ between 6L and 1L. For BN on SiO$_{2}$, surface effects might hide a strong quantum yield increase as already pointed out in the first paper of Mak et al. on MoS$_{2}$ monolayer \cite{Mak2010} and later further investigated by Buscema et al.\cite{Buscema2014} Non radiative surface states at the SiO$_{2}$/BN interface or charge transfers to excitons could for instance hinder excitonic recombinations mechanism. Alternatively, the question rises whether a strong quantum yield increase could basically occur for the BN monolayer.  Indeed, on the one hand, although some experiment led to propose a direct band gap, \cite{Watanabe2004} an indirect band gap is expected in the bulk from most  theoretical calculations. On the other hand, bulk hBN is a strongly radiative material compared to other indirect gap materials. In fact, it is clear now that the electronic structure alone is not sufficient to explain hBN luminescence properties and that unusually strong excitonic effects are involved. \\

We now turn to the thickness dependent behavior of S lines. Actually, all {\it ab initio} calculations agree to predict for the monolayer a direct gap at points K, just as for di-chalcogenides,\cite{Blase1995,Wirtz2009,Ribeiro2011,Berseneva2013} but the situation is simpler here since only two bands, the $\pi$ valence  band and the $\pi^*$ conduction band have to be considered. As a consequence the lowest excitonic state is a double degenerate state (dark triplet states are not considered here) with a large binding energy about 2 eV.\cite{Wirtz2006} Within a semi-quantitative 2D Wannier-Mott model (k.p approximation) the Bohr radius is found about 8\AA, \cite{Cao2013} which agrees with a more accurate tight-binding model.
\footnote{F. Ducastelle, 2015, unpublished}
The excitonic wave function has a typical triangular symmetry: if the hole is fixed on a nitrogen atom, the electronic density is centered on the neighboring boron atoms, just as in the bulk. \cite{Arnaud2006, Arnaud2008,Wirtz2008}

Moreover, the exciton in bulk hBN is extremely compact: the electron-hole pair extends over a few lattice parameters in the basal plane, but has an extremely limited extension over adjacent surrounding planes. That is the reason why the local structure of this exciton is not expected to be dramaticaly perturbed when stacking hBN planes. In some sense we can consider that it behaves as a Frenkel exciton along the stacking axis with a weak probability to jump from one plane to the other. In the case of hBN with the usual AA' stacking where boron atoms are above nitrogen atoms and conversely, the two types of planes are really ``optically" different and as confirmed by a simple tight-binding analysis, we expect a so-called Davydov splitting of the single layer exciton.\cite{Toyozawa2003} Hence the two 2-fold excitons found in the {\it ab initio} calculations, a dark one and a bright one. As pointed out previously, the further (Jahn-Teller or Peierls-like) splitting of these states into four states and the corresponding induced oscillator strengths are certainly related to exciton-phonon interactions but the precise mechanism is still unknown. So to summarize, a single excitonic line is expected for the monolayer which gets splitted into four lines in the bulk hBN.

In any case our experiments seem to indicate that when decreasing the number of layers we tend to the situation where we observe the expected luminescence of a single layer with a single S level. Since we have not been able to observe luminescence below 6L --- characterization of self-standing flakes is still under progress --- this is not easy to interpret. One possibility just mentioned above is that the effective thickness for luminescence processes is reduced by a dead-layer effect produced by the SiO$_{2}$ substrate interaction. Another one could be that the distortions and the symmetry lowering occurring in the bulk are no longer stable when decreasing the number of layers. Moreover, the idea that the monolayer exciton luminescence manifests itself from 100L and becomes dominant for hBN flakes below 20L could be explained by various approaches which have not been addressed for other 2D materials yet.

In conclusion, intrinsic luminescence of both bulk and exfoliated layers was identified as a function of the number of layers. Inspection of bulk crystals from several sources confirm that the intrinsic excitonic luminescence is composed of four main lines, the S lines, whereas for BN hexagonal layers it strongly depends on the number of layers and tends towards the theoretically expected behavior of a single layer due to the direct recombination of a free bright exciton. This thickness dependent behavior provides an unprecedented spectroscopic characterization of BN layers from the bulk down to a few layers. Exciton-phonons interactions would remain to study. Finally it shows that the detailed excitonic properties of bulk hBN or multilayers depend on the stacking sequences. This work confirms that h-BN displays unique properties within the family of layered semiconductors and paves the way towards future experimental and theoretical investigations of these properties .

\acknowledgments
Catherine Journet-Gautier and Berang\`ere Toury-Pierre, from LMI, are warmly acknowledged for providing one of their PDCs samples. Authors thank T. Taniguchi and K. Watanabe from NIMS for providing us a reference HPHT crystal, F. Withers from Manchester University for single, bi and tri-layer hBN samples, C. Vilar for technical help on cathodoluminescence-SEM setup, F. Fossard for helpful discussions, and Ch. Voisin for a careful reading of the manuscript. The research leading to these results has received funding from the European Union Seventh Framework Programme under grant agreement no. 604391 Graphene Flagship. We acknowledge funding by the French National Research Agency through Project No. ANR-14-CE08-0018.\\

\newpage


%

\onecolumngrid

\hrulefill

\section*{Supplementary Information}

\renewcommand{\figurename}{SFig.}
\setcounter{figure}{0}

\section*{Calibration of thickness measurement}

As pointed out in the main body of the text, due to the hydrophilic character of the SiO$_{2}$ surface, it tends to absorb water and results in a typically 1 nm-thick water layer captured between SiO$_{2}$ and BN. As for graphene, this complicates the thickness measurements of atomically-thin BN flakes by the AFM technique. Indeed we have observed that the water layer thickness is relatively inhomogeneous over the wafer, which induces a strong inaccuracy in determining the thickness of few-monolayer flakes by AFM. Thickness measurement by optical contrast (OC) imaging under visible light illumination is then preferred in this work, because it appears to be less perturbed by the presence of water. By using Differential Interference Contrast microscopy (DIC), the OC was first calibrated with AFM measurements specifically taken on BN flakes having folded parts. This ensures the step height between the folded part and the unfolded one, to be independent on the amount of water trapped between hBN and SiO$_{2}$. 

To that end, a Dimension 3100 scanning probe microscope from Brukers company equipped with commercial MPP11-100 AFM tips was used in tapping mode. The OC was measured with the same parameters (filters, exposure time) for two folded flakes deposited onto the SiO$_{2}$/Si substrate as shown in SFig.1a and SFig.1b. The values are plotted as a function of the number of layer measured by AFM in SFig.1c. We observe a linear dependence of the OC on the hBN thickness up to 20L, with approximately 1.5\% of contrast per layer in good agreement with the literature\cite{Gorbachev2011} much lower than the $\sim$\% OC of graphene.\cite{Blake2007} It is remarkable that the OC line crosses the axis at 0L, which confirms the small dependence of OC-based thickness measurements on the presence of water. Considering the small thickness of the water layer (1nm) compared to the SiO$_{2}$ thickness (90 nm) and the small refractive index difference between SiO$_{2}$ (1.45) and water (1.33) compared to hBN (1.8), the contrast measurement appears to be robust upon water contamination and allows a more accurate thickness determination than AFM for hBN thicknesses below 20L-6 nm.\\
\\

\begin{figure}[ht]
\begin{center}
\includegraphics[scale=0.9]{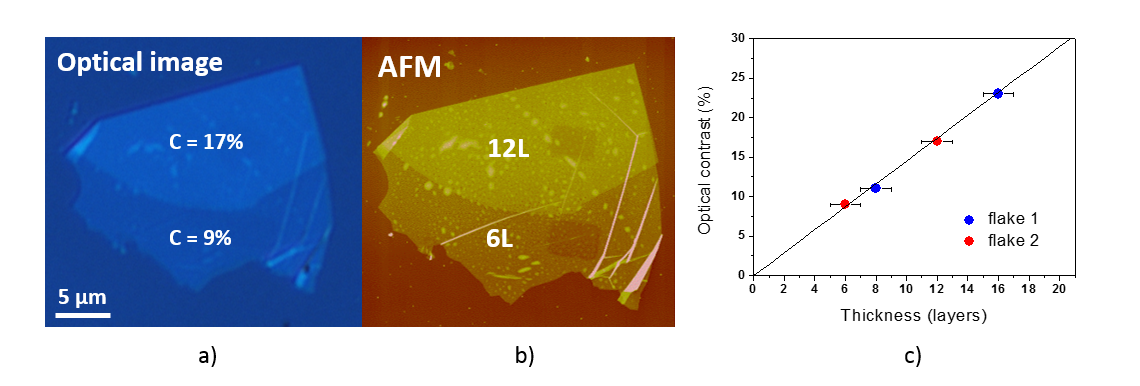}
\caption{a) DIC optical and b) AFM images of hBN folded flake. c) Calibration plot of the optical contrast (OC) as a function the thickness measured by AFM for two folded hBN flakes.}
\end{center}
\end{figure}

\section*{The low-dimensionality effect is not sample-dependent}

\begin{figure}[ht]
\begin{center}
\includegraphics[scale=0.8]{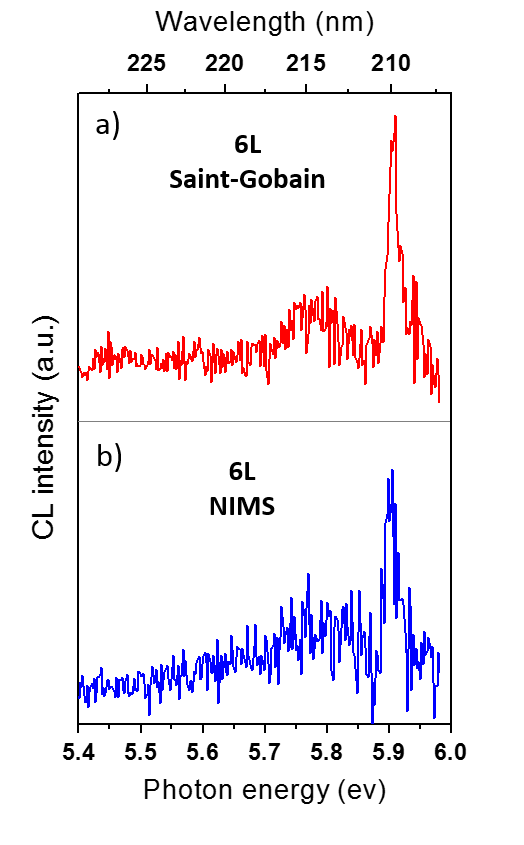}
\caption{CL spectra of 6L mechanically exfoliated hBN flakes obtained from a) Saint-Gobain powder b) NIMS sample. Spectra are corrected from the spectral response of the detection system.}
\end{center}
\end{figure}

\vfill
\newpage

\section*{Phonon replicas}
                                                                                                                                                                                                                                                                                                                                                                                                                                                                                                                                                                                                                                                                                                                                                                                                                                                                                                                                                            \begin{figure}[ht]
\begin{center}
\includegraphics[scale=0.7]{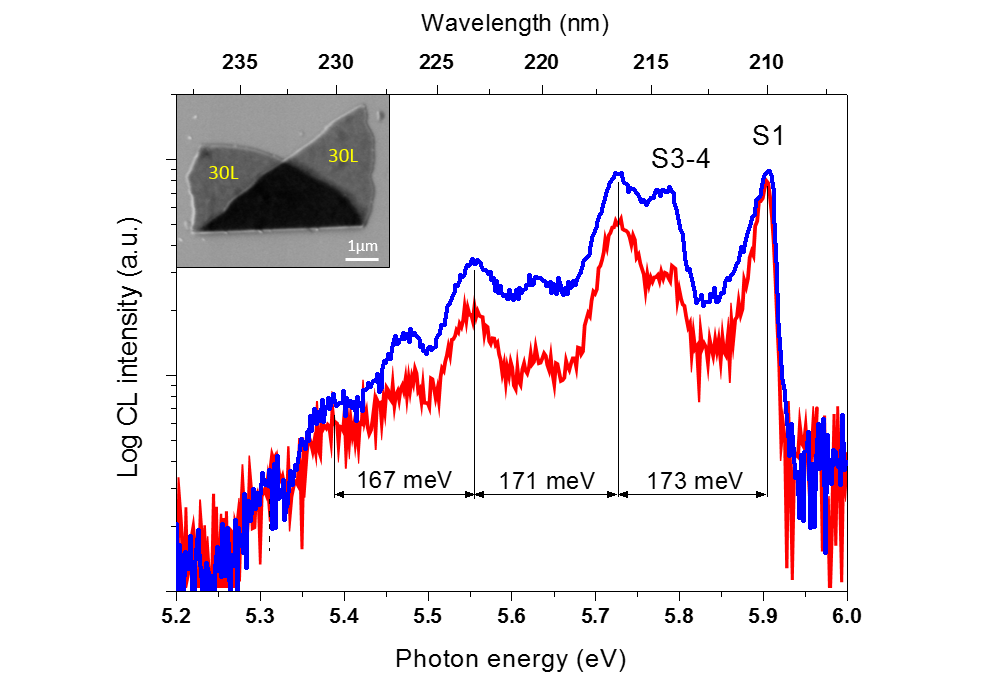}
\caption{CL spectra of hBN flakes of 24L (red) and 30L (blue, SEM image in inset) exfoliated from the Saint-Gobain powder. Note that the spectra are plotted in logarithmic scale.}
\end{center}
\end{figure}

Recorded using higher exposure time, the CL spectra of SFig.\ 3 exhibit lower energy replicas of S1 and S3-4 lines. Associated with the S1 emission at 5.903eV, we detect peaks at 5.730eV, 5.559eV and 5.392eV with decreasing intensities. The peaks are shifted by a constant energy value of $\sim$171 meV. This energy almost exactly corresponds to the 169 meV energy of the LO phonon (Raman-active) mode E$_{2g}$ of hBN. This mode is related to in-plane vibrations that is the reason why the number of hBN layers does not affect significantly its energy as shown experimentally\cite{Gorbachev2011} and theoretically.\cite{Wirtz2003} In SFig. 3, we also observe the phonon replicas of the S3-4 lines already reported in our previous study.\cite{Pierret2014}

\newpage

\section*{Relative intensities of S lines as a function of the thickness}

In SFig. 4 we plot the intensity ratio between S1 and S3 luminescence intensities for several exfoliated hBN flakes with various thicknesses. One can easily note the increasing value of the S1/S3 ratio when reducing hBN thickness. According to our interpretation the S1 line is probably related to the single sheet exciton which can be considered as a surface contribution, while the S3 line appears as a characteristic of bulk hBN. Hence, the $I_{S1}/I_{S3}$ ratio is expected to be strongly dependent on the surface/volume ratio (1/d). Then, the fit was obtained by using the expression $I_{S1}/I_{S3}=a/d^{\alpha}+b$ where $b=0.19$ is a fixed parameter whose value is systematically measured for submicrometric hBN samples. We found that the best fit is obtained for $\alpha=1.23, a=29$. Finding a value of $\alpha$ close to 1 is in a good agreement with the surface origin of the S1 peak.

\begin{figure}[ht]
\begin{center}
\includegraphics[scale=0.6]{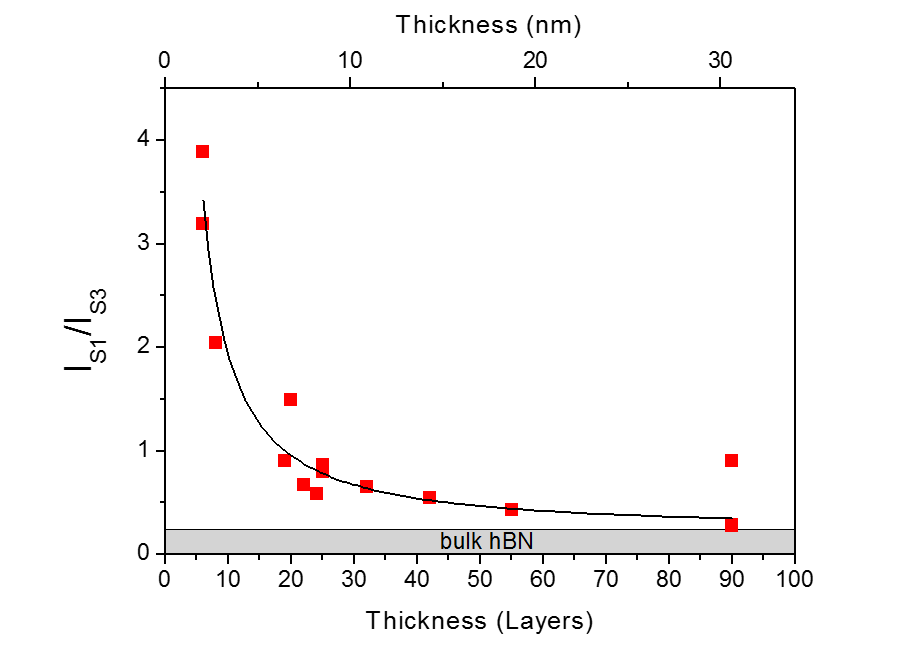}
\caption{Intensity ratio between S1 and S3 luminescence intensities as a function of the number of hBN layers. The signal amplitudes were measured for fixed energies at 5.898eV/210.2nm for the S1 emission and at 5.797eV/213.9nm for the S3 emission. The black line is a fit described in the text below.}
\end{center}
\end{figure}

\newpage

\section*{AFM characterization of thin hBN flakes}

\begin{figure}[ht!]
\begin{center}
\includegraphics[scale=0.6]{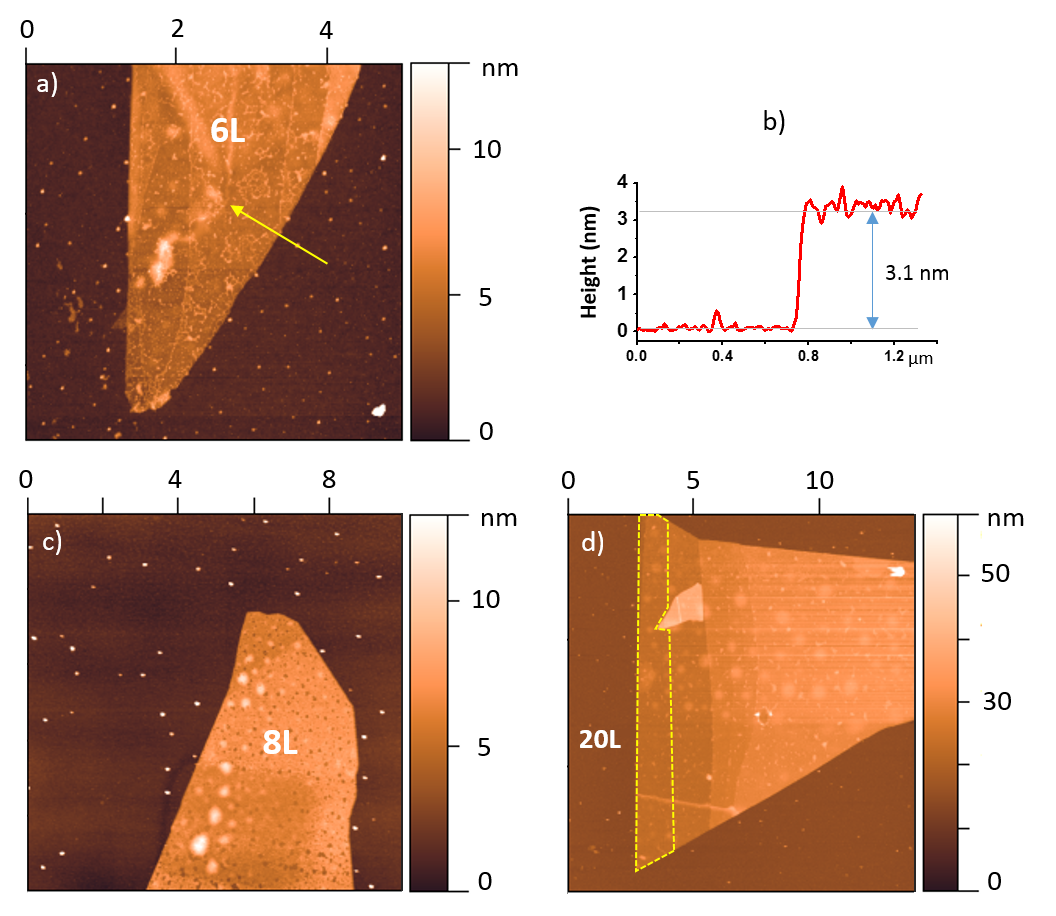}
\caption{AFM images of exfoliated hBN layers transferred onto a SiO$_{2}$/Si substrate composed of : a) 6L c) 8L and d) 20L. The image size unit is micrometer. Profile taken across the part of the 6L flake along the arrow is shown in b). The OC indicates a 6L thickness while AFM exhibits 3.1 nm height revealing a 1nm water layer thickness trapped between hBN and SiO$_{2}$.}
\end{center}
\end{figure}

\end{document}